\documentclass{iopart}
\usepackage{iopams}
\usepackage{amssymb}
\usepackage{bm}
\usepackage{graphicx}
\begin{document}

\title{Canonical separation of angular momentum of light into its orbital and spin parts}
\author{Iwo Bialynicki-Birula$^1$ and Zofia Bialynicka-Birula$^2$}
\address{$^1$ Center for Theoretical Physics, Polish Academy of Sciences,\\Al. Lotnik\'ow 32/46, 02-668 Warsaw, Poland\\ $^2$ Institute of Physics, Polish Academy of Sciences,\\Al. Lotnik\'ow 32/46, 02-668 Warsaw, Poland}
\ead{birula@cft.edu.pl}

\begin{abstract}
It is shown that the photon picture of the electromagnetic field enables one to determine unambiguously the splitting of the total angular momentum of the electromagnetic field into the orbital part and the spin part.
\end{abstract}
\submitto{Journal of Optics}

\section{Introduction}

Many authors emphasized difficulties encountered in the separation of the total angular momentum of light into its orbital and spin parts \cite{lm1}--\cite{oam}. A popular formula expressing this separation, presented in several textbooks \cite{wentzel}--\cite{merz} and used in \cite{lm1}--\cite{barnett}, has the form
\begin{eqnarray}\label{gott}
\int\!d^3r\,{\bi r}\times\left(\epsilon_0{\bi E}\times{\bi B}\right)=\int\!d^3r\,\epsilon_0 E_i({\bi r}\times{\bm\nabla})A_i+\int\!d^3r\,\epsilon_0{\bi E}\times{\bi A}.
\end{eqnarray}
This prescription is marred by a defect: the splitting is gauge dependent because it involves the vector potential ${\bi A}$. This problem has been resolved by an ad hoc postulate that the potential must be evaluated in the transverse gauge but this prescription lacks a deeper foundation. In other papers the separation has been given only for monochromatic fields or in the paraxial approximation. An additional problem that has not been resolved to the satisfaction of many authors was caused by their wish to disentangle completely the orbital and spin degrees of freedom. This is possible for massive particles but not for massless particles. The direction of the spin for all massless particles is firmly locked onto the direction of momentum: it can only be parallel or antiparallel to momentum. In other words, the helicity of massless particles---the projection of its total angular momentum on the direction of momentum--- can only take the values $\pm s$. This fact makes it impossible to rotate independently the orbital and spin degrees of freedom of photons.

The correct, gauge invariant separation of the total angular momentum into its orbital and spin parts has been proposed long time ago by Darwin \cite{darwin}. The Darwin's classic paper was cited without a comment only in \cite{jh,barnett}. In several papers from our list \cite{barnett}, \cite{li}--\cite{aml} the authors rederived his result, usually in a special case of monochromatic waves. The Darwin formula is based on the Fourier transforms of the electromagnetic field. Therefore, it does not suffer from gauge dependence. With slight changes of notation it reads:
\begin{eqnarray}\label{darwin}
\qquad\fl\int\!d^3r\,{\bi r}\times\left(\epsilon_0 {\bi E}\times{\bi B}\right) =-2\rmi\epsilon_0\int\!\frac{d^3k}{c|{\bi k}|}\left[E_i^*({\bi k})({\bi k }\times{\bm\nabla}_{\bi k})E_i({\bi k})+{\bi E}^*({\bi k})\times{\bi E}({\bi k})\right],
\end{eqnarray}
where ${\bi E}({\bi k})$ is the plane-wave component of the electric field,
\begin{eqnarray}\label{el}
{\bi E}({\bi r},t)=\int\!\frac{d^3k}{(2\pi)^{3/2}}\left[{\bi E}({\bi k})e^{-\rmi\omega_k t+\rmi{\bi k }\cdot{\bi r }}+c.c.\right],
\end{eqnarray}

In the present paper we follow in the footsteps of Darwin who wrote {\em ``The main principle of the present work is the idea that, since matter and light both possess the dual characters of particle and wave, a similar mathematical treatment should be applied to both, and that this has not been yet done as fully as should be possible''.} We show that, indeed, the wave-particle duality enables one to determine the correct separation of total angular momentum. Namely, we shall show that the Darwin separation of the total angular momentum for an arbitrary electromagnetic field into two parts follows from the photon picture of the electromagnetic field. It is in essence the separation into the part perpendicular to the photon momentum and the part parallel to the photon momentum. The first part must be identified with the orbital angular momentum whereas the second part must be identified with spin---represented by helicity. In this way, by seamlessly joining the particle and the field aspect of electromagnetism, we complete the program started by Darwin. Our analysis of the angular momentum of light starts from the quantum mechanical description of photons.

\section{Quantum mechanics of photons}

There is no consensus as to what represents the photon wave function in the coordinate representation (cf., \cite{pwf}) but there is no disagreement as to the meaning of the photon wave function in momentum space. This wave function has been introduced in the early years of quantum electrodynamics by Fock \cite{fock} and was used as a standard concept in textbooks and monographs \cite{ab}--\cite{ct}. Once we accept the existence of the photon wave function in momentum space we should define the action of various operators representing physical quantities.

In a relativistic theory---and there is no nonrelativistic theory of photons---we should first of all define the operators representing ten generators of the Poincar\'e group: the generators of translation in space (momentum ${\hat{\bm P}}$), translation in time (energy ${\hat H}$), rotation (angular momentum ${\hat{\bm J}}$), and Lorentz boosts (moment of energy ${\hat{\bm K}}$). These operators must obey the following commutation relations appropriate for the Poincar\'e group:
\numparts
\begin{eqnarray}
{}[{\hat H},{\hat P}_i]=0,\quad[{\hat H},{\hat J}_i]=0,
\quad[{\hat H},{\hat K}_i]=-\rmi\hbar c{\hat P}_i,\label{crelh}\\
{}[{\hat P}_i,{\hat P}_j]=0,
\quad[{\hat J}_i,{\hat J}_j]=\rmi\hbar\epsilon_{ijk}{\hat J}_k,
\quad[{\hat K}_i,{\hat K}_j]=-\rmi\hbar c^2\epsilon_{ijk}{\hat J}_k,\\
{}[{\hat J}_i,{\hat P}_j]=\rmi\hbar\epsilon_{ijk}{\hat P}_k,
\quad[{\hat J}_i,{\hat K}_j]=\rmi\hbar\epsilon_{ijk}{\hat K}_k,
\quad[{\hat K}_i,{\hat P}_j]=\rmi\hbar\delta_{ij}{\hat H}.\label{crelk}
\end{eqnarray}
\endnumparts
There are no problems with the construction of the generators for {\em massive} particles. The following set of operators was given long time ago by Foldy \cite{foldy}
\numparts\label{foldy}
\begin{eqnarray}
{\hat H}=E_{\bi p},\\
{\hat{\bm P}}={\bi p},\\
{\hat{\bm J}}=\rmi\hbar{\bm\nabla}_{\bi p}\times{\bi p}+{\bi S},\\
{\hat{\bm K}}={\rmi}\hbar E_{\bi p}{\bm\nabla}_{\bi p}-\frac{{\bi S}\times{\bi p}}{mc^2+E_{\bi p}},
\end{eqnarray}
\endnumparts
where ${\bm\nabla}_{\bi p}$ denotes the gradient with respect to the components of momentum and the spin vector ${\bi S}$ is built from three $(2s+1)\times(2s+1)$ matrices that obey the commutation relations of angular momentum. The matrices ${\bi S}$ act on the $(2s+1)$-component wave functions describing the states of a particle with spin $s$. In this case, the splitting of the angular momentum into its orbital and spin parts is quite obvious.

The representation of the generators of the Lorentz group for {\em massless} particles was given by Lomont and Moses \cite{lm}. We will use here a modified version of these generators \cite{bb,bb1} that exhibits its geometrical meaning. The momentum operator, by definition, acts on the wave functions in momentum representation as a multiplication by $\hbar{\bi k}$. There is no question that the operator representing the energy of the photon (the Hamiltonian) must be the modulus of the momentum vector multiplied by $c$. The complete list of generators contains also the operator of angular momentum and the boost operator,
\numparts
\begin{eqnarray}
{\hat H}=\hbar\omega_{\bi k},\label{gena}\\
{\hat{\bm P}}=\hbar{\bi k},\\
{\hat{\bm J}}=\rmi\hbar{\bi D}\times{\bi k}+\hbar{\hat\chi}{\bi n}_{\bi k},\\
{\hat{\bm K}}={\rmi}\hbar\omega_{\bi k}{\bi D},\label{gend}
\end{eqnarray}
\endnumparts
where ${\bi n}_{\bi k}={\bi k}/|{\bi k}|$, the photon helicity operator $\hat\chi$ has the eigenvalues $\pm 1$, and ${\bi D}$ stands for the covariant derivative on the light cone (${\bm\nabla}_{\bi k}=\partial/\partial{\bi k}$),
\begin{eqnarray}\label{cd}
{\bi D}={\bm\nabla}_{\bi k}-\rmi{\hat\chi}{\bm\alpha}({\bi k}).
\end{eqnarray}
These operators act on the two-component photon wave functions
\begin{eqnarray}\label{pwf}
{\bm{\mathfrak f}}({\bi k})=\left(\begin{array}{c}f_L({\bi k})\\f_R({\bi k})\end{array}\right),
\end{eqnarray}
and satisfy the commutation relations (\ref{crelh})--(\ref{crelk}) appropriate for the Poincar\'e group. The two components of the photon wave function correspond to two eigenvalues of ${\hat\chi}$,
\begin{eqnarray}\label{chi}
{\hat\chi}\left(\begin{array}{c}f_L({\bi k})\\f_R({\bi k})\end{array}\right)=
\left(\begin{array}{c}f_L({\bi k})\\-f_R({\bi k})\end{array}\right).
\end{eqnarray}
We used the indices $L$ and $R$ to denote the eigenfunctions of the helicity operator since they correspond to left-handed and right-handed circular polarization. The properties of the covariant derivative are obtained from the commutation relations for the angular momentum and they read:
\begin{eqnarray}\label{cr}
[D_i,D_j]=\rmi{\hat\chi}\epsilon_{ijl}n_l/|{\bi k}|^2.
\end{eqnarray}
These conditions determine the vector ${\bm\alpha({\bi k})}$ up to a gauge transformation
\begin{eqnarray}\label{gt}
{\bm\alpha}({\bi k})\to {\bm\alpha}({\bi k})+{{\bm\nabla}_{\bi k}}\varphi({\bi k}),
\end{eqnarray}
which is connected to the change of the phase of the wave function, in analogy to the theory of charged particles coupled to electromagnetic field. The generators (\ref{gena})--(\ref{gend}) are hermitian with respect to the following Lorentz-invariant scalar product
\begin{eqnarray}
\fl\qquad\langle{\bm{\mathfrak f}}|{\bm{\mathfrak g}}\rangle=\int\frac{d^3k}{\hbar\omega_{\bi k}}{\bm{\mathfrak f}}^\dagger({\bi k})\!\cdot\!{\bm{\mathfrak g}}({\bi k})=\int\frac{d^3k}{\hbar\omega_{\bi k}}\left[f_L^*({\bi k})g_L({\bi k})+f_R^*({\bi k})g_R({\bi k})\right].
\end{eqnarray}

\section{Electromagnetic field}

In order to solve the problem of the total angular momentum separation into two parts for the classical electromagnetic field, we shall employ the correspondence between the fundamental physical quantities (energy, momentum, and angular momentum) in photon quantum mechanics and in Maxwell theory. In quantum mechanics of photons these quantities are represented by the operators (\ref{gena})--(\ref{gend}). In Maxwell theory these quantities are given as space integrals of corresponding densities built from quadratic expressions in field vectors. A very convenient tool in this construction is a complex vector ${\bi F}$,
\begin{eqnarray}\label{rs}
{\bi F} = \sqrt{\frac{\epsilon_0}{2}}({\bi E}+ \rmi c{\bi B}),
\end{eqnarray}
that was named the Riemann-Silberstein (RS) vector in \cite{pwf}. The Maxwell equations expressed in terms of ${\bi F}$ are:
\begin{eqnarray}\label{max}
\partial_t{\bi F}({\bi r},t) = -\rmi c\nabla\times{\bi F}({\bi r},t),\quad\nabla\!\cdot\!{\bi F}({\bi r},t)=0.
\end{eqnarray}

The field energy $H$, the field momentum ${\bm P}$, the field angular momentum ${\bm J}$, and the field moment of energy ${\bm K}$ can all be constructed from the energy-momentum tensor of the electromagnetic field. These quantities expressed in terms of the RS vector are:
\numparts
\begin{eqnarray}
H=\frac{1}{2}\!\int\!d^3r\left[\epsilon_0{\bm E}^2+{\bm B}^2/\mu_0\right]=\int\!d^3r\,{\bi F}^*\!\cdot\!{\bi F},\label{genema}\\
P=\int\!d^3r\left[\epsilon_0{\bm E}\times{\bm B}\right]=\frac{1}{2\rmi}\int\!d^3r{\bi F}^*\times{\bi F},\\
{\bm J}=\int\!d^3r\,{\bi r}\times\left[\epsilon_0{\bm E}({\bm r})\times{\bm B}({\bm r})\right]=\frac{1}{2\rmi}\int\!d^3r\,{\bi r}\times\left({\bi F}^*\times{\bi F}\right),\\
{\bm K}=\frac{1}{2}\!\int\!d^3r\,{\bi r}\left[\epsilon_0{\bm E}^2+{\bm B}^2/\mu_0\right]=\int\!d^3r\,{\bi r}\,\left({\bi F}^*\!\cdot\!{\bi F}\right).\label{genemd}
\end{eqnarray}
\endnumparts
These quantities, like their counterparts in photon quantum mechanics (\ref{gena})--(\ref{gend}) serve as the generators of Poincar\'e transformations of the electromagnetic field. They have analogous algebraic properties of the Poincar\'e group (\ref{crelh})--(\ref{crelk}), with quantum commutators replaced by Poisson brackets, $[a,b]/\rmi\hbar\to \{a,b\}$, (cf., for example \cite{bb}).

All solutions of Maxwell equations in vacuum can be decomposed into plane waves with positive and negative frequencies. This decomposition gives the following Fourier representation of ${\bi F}({\bi r},t)$:
\begin{equation}\label{four}
\fl\qquad{\bi F}({\bi r},t)=\sqrt{N}\int\!\frac{d^3k}{(2\pi)^{3/2}}{\bi e}({\bi k})\left[f_L({\bi k})e^{-\rmi\omega_{\bi k} t+\rmi{\bi k}\cdot{\bi r}}+f_R^*({\bi k})e^{\rmi\omega_{\bi k} t-\rmi{\bi k}\cdot{\bi r}}\right],
\end{equation}
where the complex polarization vector ${\bi e}({\bi k})=[{\bi l}_1({\bi k})+\rmi{\bi l}_2({\bi k})]/\sqrt{2}$ has the following properties:
\numparts
\begin{eqnarray}
c{\bi k}\times{\bi e}({\bi k})=-\rmi\,\omega_{\bi k}\,{\bi e}({\bi k}),\label{pola}\\
{\bi e}({\bi k})\!\cdot\!{\bi e}({\bi k})=0,\\
{\bi e}^*({\bi k})\!\cdot\!{\bi e}({\bi k})=1,\\
{\bi e}^*({\bi k})\times{\bi e}({\bi k})=\rmi{\bi n}_{\bi k},\label{polb}\\
{\bi e}^*({\bi k})\!\cdot\!{\bi e}(-{\bi k})=0,\\
{\bi e}({\bi k})\times{\bi e}({\bi k})=0,\label{polc}\\
e_i^*({\bi k})e_j({\bi k})=\frac{1}{2}\left(\delta_{ij}+\rmi\varepsilon_{ijl}\frac{k_l}{|\bi k|}\right).\label{pold}
\end{eqnarray}
\endnumparts
The identification of the Fourier coefficients with the components of the photon wave function in the formula (\ref{four}) will be justified in the next section where we will unify the field picture and the photon picture. The second term in (\ref{four}) involves complex conjugation. This is dictated by the fact that the photon energy is always positive. Therefore, the time evolution of the wave function is given by the factor $\exp(-\rmi\,\omega_{\bi k} t)$. Therefore, the reversal of the sign in the exponent requires complex conjugation. We pulled out the factor $\sqrt{N}$ to assure the normalization of ${\bm{\mathfrak f}}$.

\section{Separation of angular momentum}

We shall combine now the field picture and the photon picture to obtain the decomposition of the total angular momentum of the field. To this end, we substitute the Fourier representation of the field into the formulas (\ref{genema})--(\ref{genemd}).
\numparts
\begin{eqnarray}
H=N\int\frac{d^3k}{\hbar\omega_{\bi k}}{\bm{\mathfrak f}}^\dagger({\bi k})\!\cdot\!\hbar\omega_{\bi k}{\bm{\mathfrak f}}({\bi k}),\label{genfa}\\
{\bm P}=N\int\frac{d^3k}{\hbar\omega_{\bi k}}{\bm{\mathfrak f}}^\dagger({\bi k})\!\cdot\!\hbar{\bi k}{\bm{\mathfrak f}}({\bi k}),\\
{\bm J}=N\int\frac{d^3k}{\hbar\omega_{\bi k}}{\bm{\mathfrak f}}^\dagger({\bi k})\!\cdot\!\left[\rmi\hbar{\bi D}\times{\bi k}+\hbar{\hat\chi}{\bi n}_{\bi k}\right]{\bm{\mathfrak f}}({\bi k}),\label{genfc}\\
{\bm K}=N\int\frac{d^3k}{\hbar\omega_{\bi k}}{\bm{\mathfrak f}}^\dagger({\bi k})\!\cdot\!{\rmi}\hbar\omega_{\bi k}{\bi D}{\bm{\mathfrak f}}({\bi k}).\label{genfd}
\end{eqnarray}
\endnumparts
Note, that the resulting expressions have the form of quantum mechanical expectation values
\numparts
\begin{eqnarray}
H=N\langle{\bm{\mathfrak f}}|{\hat H}|{\bm{\mathfrak f}}\rangle,\label{genf1a}\\
{\bm P}=N\langle{\bm{\mathfrak f}}|{\hat{\bm P}}|{\bm{\mathfrak f}}\rangle,\\
{\bm J}=N\langle{\bm{\mathfrak f}}|{\hat{\bm J}}|{\bm{\mathfrak f}}\rangle,\\
{\bm K}=N\langle{\bm{\mathfrak f}}|{\hat{\bm K}}|{\bm{\mathfrak f}}\rangle.\label{genf1d}
\end{eqnarray}
\endnumparts
These formulas exhibit a perfect agreement between the results obtained from the particle picture and from the field picture, as Darwin had anticipated. Every value calculated for the total electromagnetic field is a product of of the quantum mechanical average value per one photon, multiplied by $N$. That means that the normalization factor $N$ is the total number of photons.
We may now unambiguously split the total angular momentum of the electromagnetic field (\ref{genfc}) into two parts. The vector ${\bm J}_o$ whose integrand is perpendicular to the wave vector is the orbital part and the vector ${\bm J}_s$ with integrand parallel to the wave vector is the spin part represented by helicity,
\numparts
\begin{eqnarray}\label{orbsp}
\fl\qquad{\bm J}_o=N\int\frac{d^3k}{\hbar\omega_{\bi k}}{\bm{\mathfrak f}}^\dagger({\bi k})\!\cdot\!\left[\rmi\hbar{\bi D}\times{\bi k}\right]{\bm{\mathfrak f}}({\bi k}),\\
\fl\qquad{\bm J}_s=N\int\frac{d^3k}{\hbar\omega_{\bi k}}{\bm{\mathfrak f}}^\dagger({\bi k})\!\cdot\!\hbar{\hat\chi}{\bi n}_{\bi k}{\bm{\mathfrak f}}({\bi k})=N\int\frac{d^3k}{\omega_{\bi k}}{\bi n}_{\bi k}\left[|f_L({\bi k})|^2-|f_R({\bi k})|^2\right],
\end{eqnarray}
\endnumparts
The final step of our analysis is the proof that the expressions for ${\bm J}_o$ and ${\bm J}_s$ coincide with those obtained by Darwin. To this end, we employ the relation between ${\bi E}({\bi k})$ and ${\bm{\mathfrak f}}({\bi k})$ that follows from the formulas (\ref{el}) and (\ref{four})
\begin{eqnarray}\label{rel}
{\bi E}({\bi k})=\sqrt{\frac{N}{2\epsilon_0}}\left[{\bi e}({\bi k})f_L({\bi k})+{\bi e}^*({\bi k})f_R({\bi k})\right].
\end{eqnarray}
Upon substituting this relation into the second term in (\ref{darwin}), with the use of the properties of the polarization vectors (\ref{polb}) and (\ref{polc}), we obtain
\begin{eqnarray}\label{darwin1}
\fl\qquad-2\rmi\epsilon_0\int\!\frac{d^3k}{c|{\bi k}|}{\bi E}^*({\bi k})\times{\bi E}({\bi k})\nonumber\\
\fl\qquad=-\rmi N\int\!\frac{d^3k}{c|{\bi k}|}\left[{\bi e}^*({\bi k})\times{\bi e}({\bi k})|f_L({\bi k})|^2+{\bi e}({\bi k})\times{\bi e}^*({\bi k})|f_R({\bi k})|^2\right]={\bm J}_s.
\end{eqnarray}
In the same way we may establish the equality of the orbital part in the Darwin form and in quantum mechanics of photons. Note that the separation of the total angular momentum into its orbital and spin parts is conserved in time since both parts are separately time independent.

As an illustration, we consider the Bessel beam characterized by the frequency $c|{\bi k}|$, the $z$-component of the total angular momentum $m$, the component $k_z$ of the wave vector in the $z$-direction, and the helicity $\pm 1$. In this case, the Darwin vector (\ref{rel}) (up to a normalization factor) as given in \cite{bb2} has the form
\begin{eqnarray}\label{bes}
\fl\qquad{\bi E}_{k,m,k_z}(k',\phi,k'_z)=\left(\begin{array}{c}-(k_z/k)\cos\phi\pm\rmi\sin\phi\\
-(k_z/k)\sin\phi\mp\rmi\cos\phi\\(k_\perp/k)^2\end{array}\right)e^{\rmi m\phi}\delta(k_\perp-k'_\perp)\delta(k_z-k'_z).
\end{eqnarray}
Since the Bessel beam has an infinite extension in space, both parts of the total angular momentum are infinite. However, their ratio is well defined. For the components in the beam direction, the ratio of the orbital to spin parts equals to $mk/k_z\mp 1$.

In order to express ${\bm J}_o$ and ${\bm J}_s$ as integrals in coordinate space we have to invert the Fourier transformation in (\ref{el}) for $t=0$ as follows:
\begin{eqnarray}\label{inv}
{\bi E}({\bi k})=\int\!\frac{d^3r}{2(2\pi)^{3/2}}e^{-\rmi{\bi k }\cdot{\bi r}}\left[{\bi E}({\bi r})+\frac{\rmi c}{|{\bi k}|}{\bi\nabla}\times{\bi B}({\bi r})\right],
\end{eqnarray}
where we made use of Maxwell equations. Inserting this formula into the Darwin expression for the spin part ${\bm J}_s$, after the integration over ${\bi k }$, we obtain
\begin{eqnarray}\label{darwin2}
{\bm J}_s=\epsilon_0\int\!d^3r\int\!\frac{d^3r'}{4\pi}{\bi E}({\bi r})\times\frac{{\bi\nabla}'\times{\bi B}({\bi r}')}{|{\bi r}-{\bi r'}|},
\end{eqnarray}
where we used the formula
\begin{eqnarray}\label{formula}
\int\!\frac{d^3k}{(2\pi)^3}\frac{e^{\rmi{\bi k }\cdot{\bi r }}}{|{\bi k}|^2}=\frac{1}{4\pi|{\bi r}|}.
\end{eqnarray}
This gauge-invariant integral representation of ${\bm J}_s$ becomes equal to the last term in (\ref{gott}) if the vector potential is identified with the following integral:
\begin{eqnarray}\label{vpot}
{\bi A}({\bi r})=\int\!\frac{d^3r'}{4\pi}\frac{{\bi\nabla}'\times{\bi B}({\bi r'})}{|{\bi r}-{\bi r'}|}.
\end{eqnarray}
This representation of the vector potential is valid in the transverse gauge, as has been anticipated. Note that the seemingly local form of the formula (\ref{gott}) is misleading because the gauge invariant vector potential is a {\em nonlocal} function of the magnetic field.

\section{Conclusions}

We have shown that the separation of the total angular momentum of the electromagnetic field into its orbital and spin parts dictated by quantum mechanics of photons reproduces the results derived from the properties of Maxwell fields by Darwin. This separation, when expressed in the form of coordinate-space integrals, coincides with the results derived heuristically by many authors, provided the vector potential is related to the magnetic field by the integral formula (\ref{vpot}). In contrast to energy, momentum, and the total angular momentum of the electromagnetic field, the orbital angular momentum and the spin parts cannot be expressed as integrals of local densities: they are intrinsically nonlocal objects.

\section*{Acknowledgments}

This research was partly supported by a grant from the the Polish Ministry of Science and Higher Education for the years 2010--2012.


\section*{References}
\begin{thebibliography}{99}
\bibitem{lm1} Lenstra D and Mandel L 1982 Angular momentum of the quantized electromagnetic field with periodic boundary conditions {\em Phys. Rev.} {\bf 26} 3428
\bibitem{jh} J\`aregui R and Hacyan S 2005 Quantum-mechanical properties of Bessel beams  {\em Phys. Rev. A} {\bf 71} 033411
\bibitem{hj} Hacyan S and J\`aregui R 2006 A relativistic study of Bessel beams {\em J. Phys. B: At. Mol. Opt. Phys.} {\bf 39} 1669
\bibitem{cpb} Calvo G F, Pic\`on A and Bagan E 2006 Quantum field theory of photons with orbital angular momentum {\em Phys. Rev. A} {\bf 73} 013805
\bibitem{barnett} Barnett S M 2010 Rotation of electromagnetic fields and the nature of optical angular momentum {\em J. Mod. Opt.} {\bf 57} 1339
\bibitem{absw} Allen L, Beijersbergen M W, Spreeuw R J C and Woerdman J P 1992 Orbital momentum of light and the transformation of Laguerre-Gaussian laser modes {\em Phys.    Rev. A} {\bf 45} 8185
\bibitem{en} van Enk J S and Nienhuis G 1992 Eigenfunction expansion of laser beams and orbital angular momentum of light {\em Opt. Comm.} {\bf 94} 147
\bibitem{na} Nienhuis G and Allen L 1993 Paraxial wave optics and harmonic oscillators {\em Phys. Re. A} {\bf 48} 656
\bibitem{ba} Barnett S M and Allen L 1994 Orbital angular momentum and nonparaxial light beams {\em Opt. Comm.} {\bf 110} 670
\bibitem{berry1} Berry M 1998 Paraxial beams of spinning light {\em Singular Optics} Eds. M S Soskin and M V Vasnietsov, SPIE 3487 6
\bibitem{barnett0} Barnett S M 2002 Optical angular momentum flux {\em J. Opt. B: Quantum Semiclassical Opt.} {\bf 4} S7
\bibitem{nienhuis} Nienhuis G 2006 Polychromatic and rotating beams of light {\em J. Phys. B: At. Mol. Opt. Phys.} {\bf 39} S529
\bibitem{berry} Berry M V 2009 Optical currents {J. Opt. A: Pure Appl. Opt.} {\bf 11} 094001
\bibitem{mazilu} Mazilu M 2009 Spin and angular momentum operators and their conservation {J. Opt. A: Pure Appl. Opt.} {\bf 11} 094005
\bibitem{li} Chun-Fang Li 2009 Spin and orbital angular momentum of a class of nonparaxial light beams having a globally defined polarization {\em Phys. Rev. A} {\bf 80} 063814
\bibitem{alml} Aiello A, Lindlein N, Marquardt C and Leuchs G 2009 Transverse angular momentum and geometric spin Hall effect of light {\em Phys. Rev. Lett.} {\bf 103} 100401
\bibitem{aml} Aiello A, Marquardt C and Leuchs G 2009 Transverse angular momentum of photons {\em Phys. Rev. A} {\bf 81} 053838
\bibitem{oam} Allen L, Barnett S M and Padget M J 2003 {\em Optical Angular Momentum} (Bristol: Institute of Physics Publishing) This is a collection of papers with introductory material by the editors
\bibitem{wentzel} Wentzel G 1949 {\em Quantum Theory of Fields} (Interscience: New York) Ch. 4
\bibitem{gott} Gottfried K 1966 {\em Quantum Mechanics} (New York: Benjamin) Ch. 8
\bibitem{merz} Merzbacher E 1970 {\em Quantum Mechanics} (Wiley: New York) Ch. 22
\bibitem{darwin} Darwin C G 1932 Notes on the theory of radiation {\em Proc. Roy. Soc. London} {\bf 136} 36
\bibitem{pwf} Bialynicki-Birula I 1996 Photon wave function {\em Progress in Optics} {\bf 36} ed E Wolf (Amsterdam: Elsevier) (ArXiv: quant-ph/0508202)
\bibitem{fock} Fock V 1934 Zur Quantenelektrodynamik {\em Phys. Zeit. der Sowjetunion} {\bf 6} 425
\bibitem{ab} Akhiezer A I and Berestetski V B 1965 {\em Quantum Electrodynamics} (New York: Interscience) Ch.~1
\bibitem{ss} Schweber S S 1961 {\em An Introduction to Relativistic Quantum Field Theory} (Evanston: Row, Peterson and Co) Ch. 9
\bibitem{bb} Bialynicki-Birula I and Bialynicka-Birula Z 1975 {\em Quantum Electrodynamics} (Oxford: Pergamon) Ch. 9
\bibitem{pauli} Pauli W 1980 {\em General Principles of Quantum Mechanics} (Berlin: Springer) Ch. 25
\bibitem{ct} Cohen-Tannoudji C, Dupont-Roc J and Grynberg G 1989 {\em Photons and Atoms: Introduction to Quantum Electrodynamics} (New York: Wiley) Ch. 1
\bibitem{foldy} Foldy L L 1956 Synthesis of covariant particle equations {\em Phys. Rev.} {\bf 102}, 568. This form of the generators is taken from \cite{lm}.
\bibitem{lm} Lomont J S and Moses H E 1962 Simple realizations of the infinitesimal generators of the proper orthochronous inhomogeneous Lorentz group for mass zero {\em J. Math. Phys.} {\bf 3}, 405
\bibitem{bb1} Bialynicki-Birula I and Bialynicka-Birula Z 1987 Berry's phase in the relativistic theory of spinning particles, {\em Phys. Rev. D} {\bf 35} 2383
\bibitem{bb2} Bialynicki-Birula I and Bialynicka-Birula Z 2006 Beams of electromagnetic radiation carrying angular momentum: The Riemann-Silberstein vector and the classical-quantum correspondence {\em Opt. Comm.} {\bf 264} 342 (ArXiv: quant-ph/05011011)
\end{thebibliography}
\end{document}